# Observation of intrinsic chiral bound states in the continuum

Yang Chen[1,2†], Huachun Deng[3†], Xinbo Sha[3†], Weijin Chen[1†], Ruize Wang[2], Yuhang Chen[2], Dong Wu[2], Jiaru Chu[2], Yuri S. Kivshar[4], Shumin Xiao[3,5]*, Cheng-Wei Qiu[1]*

[1]Department of Electrical and Computer Engineering, National University of Singapore, 117583 Singapore, Singapore

[2]Chinese Academy of Sciences Key Laboratory of Mechanical Behavior and Design of Materials, Department of Precision Machinery and Precision Instrumentation, University of Science and Technology of China, 230027 Hefei, China

[3]Ministry of Industry and Information Technology Key Lab of Micro-Nano Optoelectronic Information System, Harbin Institute of Technology, Shenzhen, 518055, P. R. China

[4]Nonlinear Physics Center, Research School of Physics, Australian National University, Canberra, ACT2601, Australia

[5]Pengcheng Laboratory, Shenzhen, 518055, P. R. China

*†These authors contributed equally: Yang Chen, Huachun Deng, Xinbo Sha and Weijin Chen*

*Email: shumin.xiao@hit.edu.cn, chengwei.qiu@nus.edu.sg;*

**Summary paragraph**

Photons with spin angular momentum possess intrinsic chirality which underpins many phenomena including nonlinear optics[1], quantum optics[2], topological photonics[3] and chiroptics[4]. Intrinsic chirality is weak in natural materials, and recent theoretical proposals[5-7] aimed to enlarge circular dichroism by resonant metasurfaces supporting bound states in the continuum that enhance substantially chiral light-matter interaction. Those insightful works resort to three-dimensional sophisticated geometries, which are too challenging to be realized for optical frequencies[8]. Therefore, most of the experimental attempts[9-11] showing strong circular dichroism rely on false/extrinsic chirality by employing either oblique incidence[9, 10] or structural anisotropy[11]. Here, we report on the experimental realization of true/intrinsic chiral response with resonant metasurfaces where the engineered slant geometry breaks both in-plane and out-of-plane symmetries. Our result marks the first observation of *intrinsic chiral bound states in the continuum* with near-unity chiral dichroism of 0.93 and record-high quality factor exceeding 2663 for visible frequencies. Our chiral metasurfaces promise a plethora of applications in chiral light sources and detectors, chiral sensing, valleytronics and asymmetric photocatalysis.

**Main text**

Chirality, a fundamental trait of nature, refers to the geometric attribute of objects that lack mirror-reflection symmetry. To evaluate how chiral an object is, electromagnetic chirality, with the manifestation of circular dichroism (CD), is conventionally adopted based on the differential interactions between the object and electromagnetic fields of different handedness[12, 13]. However, it is found that planar structures with out-of-plane mirror symmetry, that are not supposed to be chiral, can still demonstrate strong CD signals through the introduction of structural anisotropy[14] or oblique incidence[15, 16]. In these cases, the amplitude of CD cannot measure the "*true chirality*" or "*intrinsic chirality*" of an object, but it is originated from anisotropy-induced polarization conversion or chiral configurations of the experimental setup, which are usually called "*false chirality*" or "*extrinsic chirality*"[15-18]. Although false chirality may yield similar CD signals as its counterpart of true chirality, its applications in a range of important fields, such as chiral emission and polarized photodetection, are significantly limited.

Apart from intrinsic chirality, another key parameter for enhancing the strength of chiral light-matter interaction is the quality ($Q$) factor of the associated resonance. Due to the potential applications in chiral emission, chiral sensing and enantiomer separation, high-$Q$ resonances with large intrinsic chirality have long been pursued but remain unexplored. Chiral metamaterials/metasurfaces can produce strong chiroptical responses[19-21], but their achieved $Q$-factors are still low, typically less than 200, due to the large radiative and nonradiative losses. Recently, the physics of bound state in the continuum (BIC) has been employed in photonics to achieve and engineer high-$Q$ resonances[22-25]. When BIC acquires intrinsic chirality, the resulting chiral BIC can simultaneously generate high $Q$-factors and strong CDs without involving false chirality. As pointed out by previous theoretical works, the key to enabling chiral BIC is to break all the mirror symmetries of the structure[5-7], which, however, has hindered its experimental realization. We have witnessed numerous approaches to break either the in-plane[24, 26] or the out-of-plane[27] mirror symmetry, but the remaining symmetry planes still prevent the generation of intrinsic chiral BIC. The measured high-$Q$ CD resonances are inevitably attributed to the false chirality of oblique incidence[9, 10] or polarization conversion[11].

Here, we report the first optical realization of intrinsic/true chiral BIC based on a new paradigm of slant-perturbation metasurfaces. The metasurface is composed of a square array of slanted trapezoid nanoholes in a $TiO_2$ film, which is placed on a glass substrate and covered with PMMA (Fig. 1a). This structure is evolved from vertical square nanoholes by introducing two types of perturbations, an in-plane deformation angle $\alpha$ and an out-of-plane slant angle $\varphi$, so that all the mirror symmetries are broken. A series of Bloch modes are supported by the metasurface (Fig. 1b), whose mode profiles are shown in Supplementary Fig. 1. Without loss of generality, we first consider the fundamental $TM_1$ mode. When no perturbations are involved ($\alpha = 0$, $\varphi = 0$), it supports a symmetry-protected BIC at the $\Gamma$ point of Brillouin Zone because of the $C_2^z$ symmetry of the structure. Due to time-reversal symmetry, the electromagnetic near fields are always linearly polarized for BIC and their distributions cancel each other to prohibit far-field radiation (see Supplementary Section 3).

Once an in-plane geometric perturbation is introduced to break the $C_2^z$ symmetry, e.g., the square nanohole is cut into a trapezoid ($\alpha \neq 0$, $\varphi = 0$), the BIC is evolved to a quasi-BIC possessing circular polarizations in the near field, whose chirality can be evaluated by optical chirality density[28, 29]: $OCD = -\frac{1}{2}\omega \text{Im}[\boldsymbol{D} \cdot \boldsymbol{B}^*]$. Since $OCD$ is a parity-odd scalar[30], the existence of a mirror symmetry forces it to have opposite values on the two sides of the mirror as shown in Fig. 1c. In the far field, the Stokes parameter $S_3$ of radiation is related to the optical chirality flux $\mathcal{F}$ by the equation: $S_3 = \frac{c}{\omega \mathcal{S}} \int_V Re(\nabla \cdot \mathcal{F}) dv$, where $\mathcal{S}$ is the power flux and $\mathcal{F}$ is defined as: $\mathcal{F} = \frac{1}{4}[\boldsymbol{E} \times (\nabla \times \boldsymbol{H}^*) - \boldsymbol{H}^* \times (\nabla \times \boldsymbol{E})]$. In analogy to Poynting's theorem, optical chirality is also bounded by the conservation law. Thus, optical chirality flux $\mathcal{F}$ is directly related to the near-field $OCD$ of the associated resonance by the equation:

$$-2\omega \int_V OCD dv + \int_V Re(\nabla \cdot \mathcal{F}) dv = 0, \qquad (1)$$

Here, the antisymmetric $OCD$ distributions cancel each other in the near field of the metasurface and hence generate no chiral flux in the far field. This is protected by out-of-plane mirror symmetry and is immune from in-plane geometries. One of the most convenient methods to break the out-of-plane mirror symmetry is to slant the nanohole in $x$-direction. Then the variation of $OCD$ is written as: $\Delta OCD = (\Delta \varepsilon / \varepsilon) \cdot OCD$, where $(\Delta \varepsilon / \varepsilon)$ denotes the change of permittivity divided by its original value. As highlighted in Fig. 1c (see the middle panel), $\Delta \varepsilon$ and $OCD$ have opposite signs in all perturbed areas and hence the volume-integrated $\Delta OCD$ is negative. The unbalanced $OCD$ distributions in the near field of the slant-perturbation metasurface will induce non-zero optical chirality flux in the far field, corresponding to circularly polarized radiation (Fig. 1c).

The origin of slant-induced chirality can also be analyzed by examining the near-field electromagnetic distributions at the central *x-y* plane. As shown in Fig. 1d, when no slant perturbation is introduced, the magnetic fields of quasi-BIC are predominantly *x*-polarized while the electric fields are out-of-plane. According to the generalized theory of chiroptics[12, 15], the optical chirality of an object in the dipole approximation is controlled by the dot product $\boldsymbol{p}_\perp \cdot \boldsymbol{m}_\perp$, where $\boldsymbol{p}_\perp$ and $\boldsymbol{m}_\perp$ are the projections of the associated electric dipole *p* and magnetic dipole *m* on the plane perpendicular to the *k*-vector of incident light. Here, *p* is parallel to *k*, resulting in no optical chirality. To break such parallel configuration, the nanohole can be slanted towards negative *x*-direction so that the associated electric field vectors are tilted towards the same direction, while the magnetic fields approximately remain *x*-polarized (Fig. 1d), leading to non-zero $\boldsymbol{p}_\perp \cdot \boldsymbol{m}_\perp$ and optical chirality.

In Fig. 2a, we calculate the unbalanced *OCD* that is equal to $\int_V OCD dv$ as a function of the slant angle for different quasi-BICs. It can now be seen that an optimal slant angle exists for attaining the maximal unbalanced *OCD*, where the largest degree of circular polarization of far-field radiation is anticipated. Besides, we can see that the amplitude of unbalanced *OCD* cannot reach unity through the slant operation alone for $TM_2$ and $TE_2$ modes. For $TE_1$ mode, the slant angle needs to be larger in order to obtain large unbalanced OCD, and inevitably leads to a much smaller *Q*-factor. Thus, $TM_1$ mode is found to be the best candidate for achieving intrinsic chiral BIC with large *Q*-factor.

The evolution of the momentum-space eigenpolarization map of $TM_1$ along with geometric perturbations is presented in Fig. 2b. For the unperturbed case, BIC is manifested by an at-$\Gamma$ V point in the map to represent a polarization singularity. Once a nonzero $\varphi$ is induced, the integer-charged V point is decomposed into a pair of half-charged C points distributed symmetrically on the two sides of $\Gamma$ point, where the C+ and C- points possess right- and left-handed circular polarization (RCP and LCP) respectively. Further, if a nonzero $\alpha$ is introduced as well, the polarization map as a whole is moved in the same direction of structural inclination. For a proper combination of $\alpha$ and $\varphi$, e.g., $\alpha = 0.12$ and $\varphi = 0.1$, the C- point is shifted further to the left, while the C+ point can be located right at the $\Gamma$ point, leading to the achievement of intrinsic chiral BIC (Fig. 2b). Similarly, we can also create chiral BIC in the $TE_1$ mode (see Supplementary Fig. 3). The role played by the index-matched PMMA layer is discussed in Supplementary Section 5.

The proposed metasurface is fabricated by a modified slanted-etching system (see details in Supplementary Fig. 5). For the accurate control of small slant angles, the sample is placed on a wedged substrate and an $Al_2O_3$ screen with an aperture is laid above the sample, acting as an ion collimator. The SEM images of fabricated samples are shown in Fig. 2c and Supplementary Fig. 6. Thanks to the usage of ion collimator, the left and right sidewalls exhibit an almost identical slant angle. The angle-resolved transmission spectra of slant-perturbation metasurface ($\alpha = 0.12$, $\varphi = 0.1$) under RCP and LCP incidence are simulated in Fig. 2d. It is observed that the C+ point represented by the diminishing point in the $TM_1$ band of LCP incidence appears at normal direction, i.e., $\Gamma$ point. But for the RCP incidence case, the TM$_1$ mode can be excited at normal incidence, and the C- point is observed at the incidence angle of -0.04 rad. The experimental results agree well with simulations, where the measured C+ and C- points are present at the incidence angles of 0 and -0.044 rad (Fig. 2d). The details of the optical experimental setup are provided in Supplementary Fig. 7.

To study the evolution of C points with the slant angle, we have fabricated a serious of metasurfaces with a fixed $\alpha$ but variable $\varphi$. As retrieved from transmission spectra, the incident angles that C+ and C- points occur approximately follow linear relationships with $\varphi$, which is consistent with simulations results (Fig. 3a). Apparently, the key point for achieving chiral BIC is to cooperatively modulate $\alpha$ and $\varphi$, so that one C points is generated and then moved back to $\Gamma$ point. To reveal such inherent linkage between $\alpha$ and $\varphi$, the generalized model based on electric and magnetic dipoles shown in Fig. 1d is revisited. When the associated perturbations $\alpha$ and $\varphi$ are small, the $Q$-factor of quasi-BIC roughly scales with the inversely quadratic square of all the perturbations[24]: $Q \sim 1/(\alpha^2 + A\varphi^2)$, where $A$ expresses the different sensitivities of $Q$ to $\alpha$ and $\varphi$. Meanwhile, the amplitudes of electric dipole $p$ and magnetic dipole $m$ are proportional to the square root of the $Q$-factor: $|p| \sim Q^{1/2}$ and $|m| \sim Q^{1/2}$. Then, the intrinsic chirality of quasi-BIC, manifested by CD, can be estimated by:

$$\text{CD} \sim \boldsymbol{p}_\perp \cdot \boldsymbol{m}_\perp = |p||m|\sin(\varphi) \sim \frac{\sin(\varphi)}{\alpha^2 + A\varphi^2} \approx \frac{\varphi}{\alpha^2 + A\varphi^2}, \qquad (2)$$

As predicted by Eq. 2, if $\varphi$ is raised from zero while $\alpha$ is fixed, CD will first rapidly increase to the maximum and then gradually decrease. This is well reproduced by simulations results (Fig.

3b). The experimental results also follow a similar dependence except that the measured CDs are smaller than the simulated ones (see detailed spectra data in Supplementary Fig. 8). Such deviation is mainly attributed to the fabrication tolerance and the undesired scattering from surface roughness. Further, by calculating the derivative of CD versus $\varphi$, the condition for maximizing CD is deduced as: $\alpha = \sqrt{A} \cdot \varphi$. This offers a straightforward recipe to select a suitable set of $\varphi$ and $\alpha$ for achieving chiral BIC. The slope $\sqrt{A}$ is related to the mode profile and could take different values for different chiral BICs. For $TM_1$ mode, the slope is theoretically predicted to be 1.066 (see Supplementary Section 9), which agrees well with the simulation results (Fig. 3c). The experimental data also follow a linear relationship and the fitted slope of 1.197 is slightly deviated from the predicted one. Accordingly, as long as $\varphi$ and $\alpha$ are cooperatively decreased, the $Q$-factor of chiral BIC can be continuously boosted while maintaining a unity CD (see details in Supplementary Section 10). In our experiments, $\varphi$ and $\alpha$ are set as 0.1 and 0.12 due to fabrication capacity. We notice that the slant direction of the nanohole can also be rotated with an azimuthal angle $\theta$, whose impact is discussed in Supplementary Section 11.

Another way to raise the $Q$-factor of chiral BIC is to enlarge the metasurface size, so that both in-plane and out-of-plane leakage are suppressed[31]. We have fabricated a group of metasurface samples with different sizes. The highest $Q$-factor of 2663 is obtained for the largest sample of 200 μm, while the maximum CD is also reached to be 0.93 (Fig. 4a). Here, CD is defined as: CD = $(R_R - R_L) / (R_R + R_L)$, where $R_{R(L)}$ is the normalized reflection spectra under RCP (LCP) illumination. Their differential near-field distributions are presented in Supplementary Fig. 12. To exclude the possible impact of structural anisotropy, we have measured the normalized reflection matrix $\boldsymbol{R} = [R_{RR}, R_{RL}; R_{LR}, R_{LL}]$ in circular basis, where the notation $R_{RL}$ refers to the reflection of RCP light under LCP incidence. As shown in Fig. 4b, the cross-polarized components, $R_{RL}$ and $R_{LR}$, possess negligible intensities, suggesting the absence of polarization conversion. It is thus concluded that the observed CD signal is attributed to the intrinsic chirality of quasi-BIC. The measured transmission spectra are included in Supplementary Fig. 13. The fundamental difference between our demonstrated intrinsic chiral BIC and other BIC works relying on false chirality to generate large CDs is explicitly discussed in Supplementary Section 14.

In Fig. 4c, we have summarized the experimental $Q$-factors and CDs from some typical works about chiral metamaterials/metasurfaces[11, 14, 32-40]. These works are divided into two categories according to the origin of CD signals: one purely relies on the true chirality of associated resonance and the other also has false chirality involved. Clearly, most approaches achieving high CDs rely on false chirality effects[11, 35, 37, 38] and their $Q$-factors are still much smaller than ours. Besides, it is noted that the previous works exhibiting relatively large $Q$-factors are inevitably conducted in the infrared spectra[11, 38, 39], highlighting the great difficulty and significance in achieving intrinsic chiral BIC in the visible.

In conclusion, we have presented the first experimental observation of optical chiral BICs enabling simultaneously record-high values of $Q$-factor ($Q$ = 2663) and near-unity CD of 0.93.

We have developed a microscopic model based on the variation of local spin density to explain the origin of optical chirality. Although our chiral BIC metasurface is demonstrated in the visible, the concept is general being applicable to the infrared and longer spectra and promising future applications for chiral light sources and detectors, chiral sensing, quantum optics, and asymmetric photocatalysis.

## Acknowledgements

S.M.X. acknowledges the support from National Key Research and Development Project (Grant No. 318 2021YFA1400802). Y.C. acknowledges the support from the start-up funding of University of Science and Technology of China and the CAS Talents Program. D.W. acknowledges the support from the National Natural Science Foundation of China (No. 61927814). C.-W.Q. acknowledges financial support from the National Research Foundation,

Prime Minister's Office, Singapore under Competitive Research Program Award NRF-CRP22-2019-0006. C.-W.Q. is also supported by a grant (R-261-518-004-720| A-0005947-16-00) from Advanced Research and Technology Innovation Centre (ARTIC) in National University of Singapore.

## Author contributions

Y.C., S.M.X. and C.-W.Q. conceived the idea and designed the experiments. S.M.X. and C.-W.Q. supervised the project. Y.C. and W.J.C. conducted the simulations and theoretical analysis. H.C.D. and X.B.S. performed the experiments. Y.C., Y.K., S.M.X. and C.-W.Q analyzed the data. Y.C. drafted the paper with inputs from all authors.

## Competing financial interests

The authors declare no competing financial interest.

# Figures and figure captions

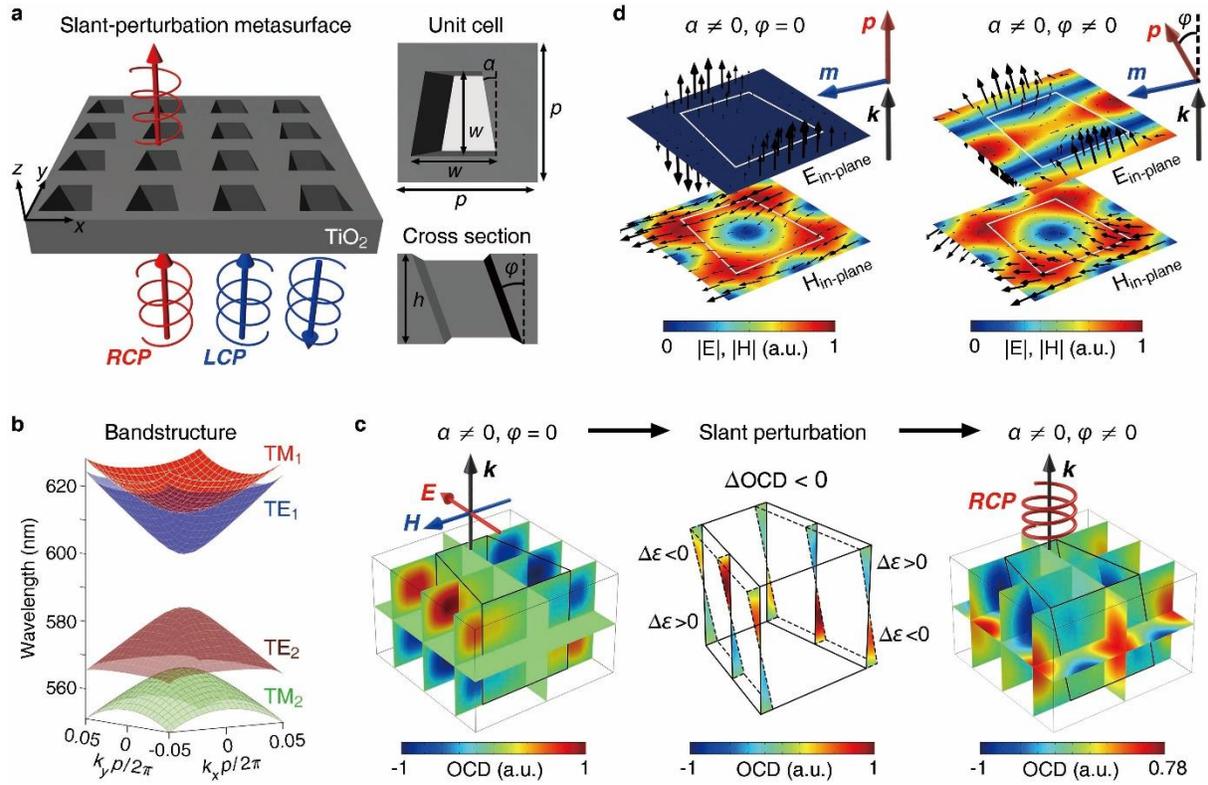

**Fig. 1 | Origin of intrinsic chirality induced by slant perturbation. a,** Schematic of the slant-perturbation metasurface to realize intrinsic chiral BIC. The geometric parameters are: $p$ = 340 nm, $w$ = 210 nm, $h$ = 220 nm. **b,** Calculated bandstructure of the metasurface with only nondegenerate modes plotted. **c,** Cross-sectional OCD (optical chirality density) distributions for the case of $\alpha \neq 0$, $\varphi = 0$ (left) and $\alpha \neq 0$, $\varphi \neq 0$ (right). OCD distributions in the slant-perturbed areas are highlighted in the middle panel with their permittivity change ($\Delta\varepsilon$) indicated. **d,** In-plane components of electric ($E_{in\text{-}plane}$) and magnetic ($H_{in\text{-}plane}$) field distributions at the central $x$-$y$ plane of the metasurface without (left) and with (right) slant perturbation, along with the configurations of corresponding electric dipole $p$ and magnetic dipole $m$.

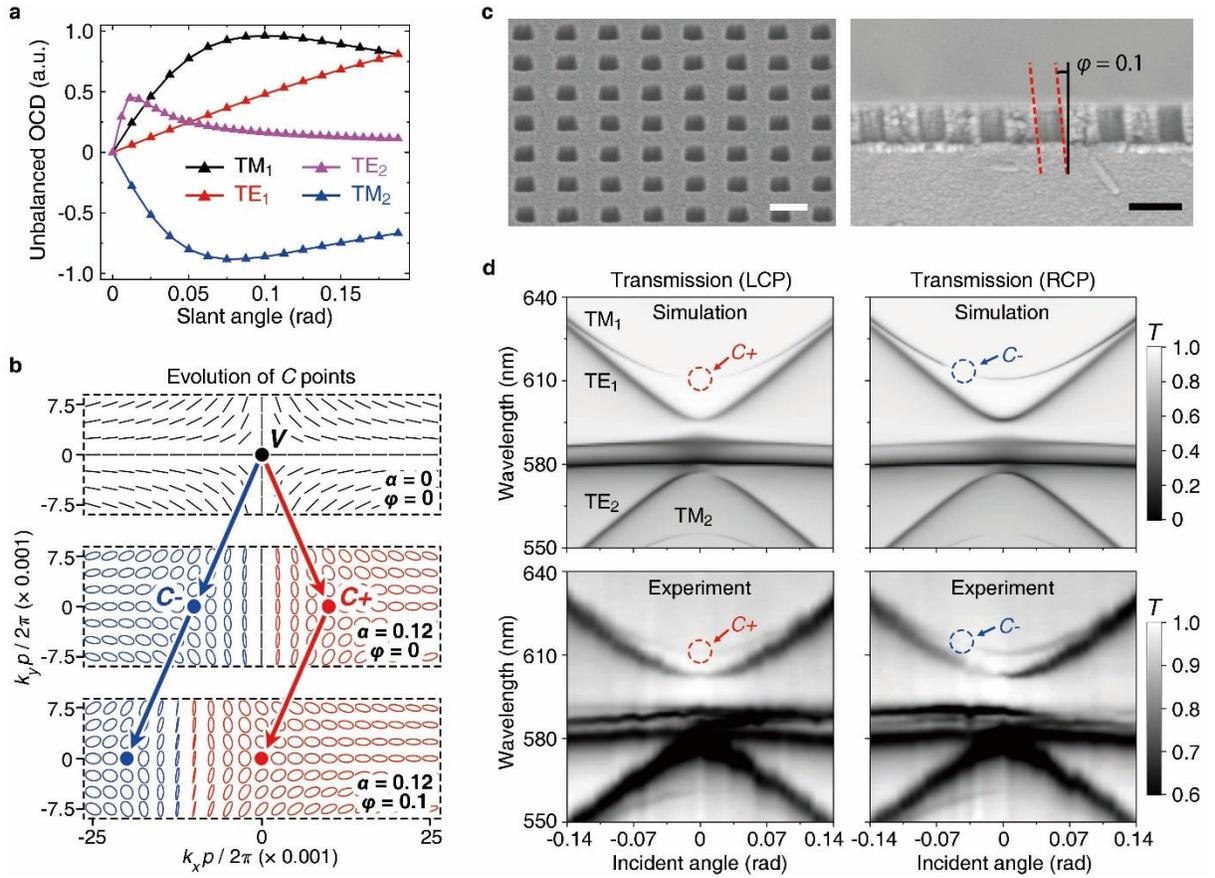

**Fig. 2 | Design, fabrication and characterization of slant-perturbation metasurfaces. a,** Unbalanced OCD integrated over the metasurface as a function of the slant angle for different quasi-BICs. **b,** Evolution of C points over $k$-space for the metasurfaces of different $\alpha$ and $\varphi$. The elliptical polarizations are represented by ellipses of red or blue colors corresponding to right- or left-handed states, while the black lines represent linear polarizations. **c,** Side-view (left) and cross-sectional (right) SEM images of a fabricated metasurface. Scale bar: 300 nm. **d,** Angle-resolved transmission spectra of the metasurface under LCP (left) and RCP (right) incidence obtained from simulations (top) and experiments (bottom). **d,** Incident angles that C+ and C- points are observed for different slant angles $\varphi$, retrieved from simulations and experiments.

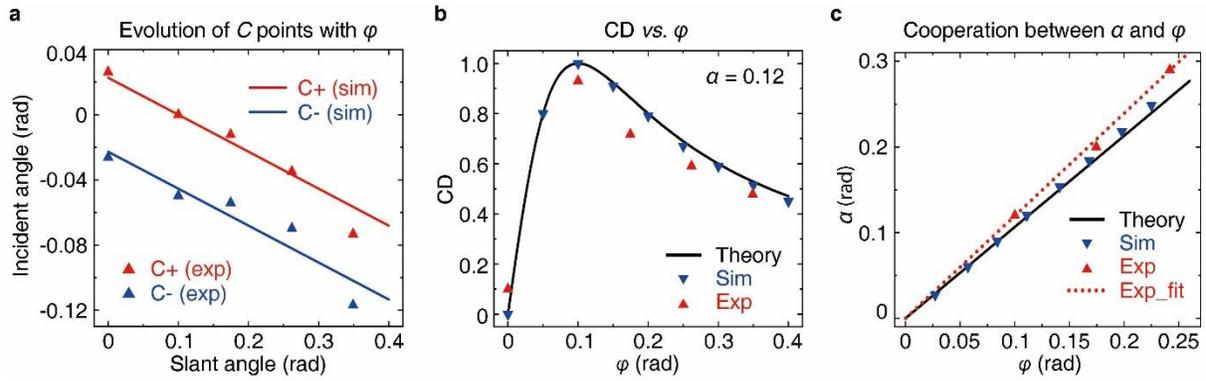

**Fig. 3 | Inherent linkage between geometric perturbations for achieving chiral BIC. a,** Incident angles that C+ and C- points are observed for different slant angles $\varphi$, retrieved from simulations and experiments. **b,** CD amplitude as a function of $\varphi$ while $\alpha$ is fixed to be 0.12. Theoretical, simulation and experimental results are included for comparison. **c,** Relation between $\varphi$ and $\alpha$ for maximizing CD. The experimental data points are fitted by a linear relation (Exp_fit).

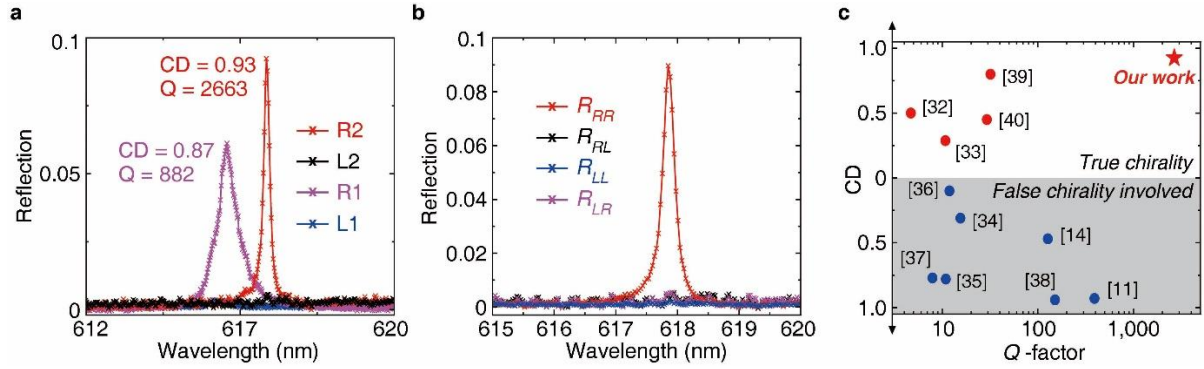

**Fig. 4 | Giant CD and *Q*-factor enabled by intrinsic chiral BIC. a,** Measured reflection spectra of the two metasurface samples of 68 μm (L1 & R1) and 200 μm sizes (L2 & R2) under LCP and RCP incidence. Their retrieved CDs and *Q*-factors are marked. **b,** Measured reflection matrix ***R*** in the circular basis for the 200 μm sample. **c,** CDs and *Q*-factors obtained from some typical experimental works as compared to our work. They are classified into two categories: true chirality and false chirality involved, according to the origin of CDs.